\documentclass[rmp,aps,preprint]{revtex4-1}

\usepackage{graphicx}

\def\inbar{\,\vrule height1.5ex width.4pt depth0pt}
\def\IR{\relax{\rm I\kern-.18em R}}
\def\IC{\relax\hbox{$\inbar\kern-.3em{\rm C}$}}

\begin{document}
\title{Universality in DAX index returns fluctuations}

\author{Rui Gon\c calves}
\affiliation{LIAAD-INESC Porto LA and Faculty of Engineering, University of Porto, R. Dr. Roberto Frias s/n, 4200-465 Porto, Portugal}
\author{Helena Ferreira}
\affiliation{LIAAD-INESC Porto LA, Portugal}
\author{Alberto Pinto}
\affiliation{LIAAD-INESC Porto LA and Department of Mathematics, Faculty of Sciences, University of Porto, Rua do Campo Alegre, 687, 4169-007, Portugal}

\begin{abstract}  
In terms of the stock exchange returns, we compute the analytic expression of the probability distributions $F_{DAX,+}$ and $F_{DAX,-}$ of the normalized positive and negative DAX (Germany) index daily returns $r(t)$. Furthermore, we  define the $\alpha$ re-scaled DAX daily index positive returns  $r(t)^\alpha$ and negative returns $(-r(t))^\alpha$ that we call, after normalization, the $\alpha$ positive fluctuations and $\alpha$ negative  fluctuations. We use the Kolmogorov-Smirnov statistical test, as a method, to find the values of  $ \alpha$ that optimize the data collapse of the histogram of the $ \alpha$  fluctuations  with the   Bramwell-Holdsworth-Pinton (BHP) probability density function. The optimal parameters that we found are $\alpha^{+}= 0.50$  and $\alpha^{-}= 0.48$. Since the BHP probability density function appears in several other dissimilar phenomena, our results reveal universality in the stock exchange markets.
\end{abstract}                                                                 

\maketitle

\section{INTRODUCTION}
\label{sec:intro}

The modeling of the time series of stock  prices is a main issue in
economics and finance and it is of a vital importance in the
management of large portfolios of stocks, see \cite{Gabaixetal03,LilloMan01} and \cite{ManStan95}.  Here we study de DAX indice.
The DAX (Deutscher Aktien IndeX, formerly Deutscher Aktien-Index (German stock index)) is a blue chip stock market index that measures the development of the 30 largest and best-performing companies on the German equities market and represents around 80 $\%$ of the market capital authorized in Germany. The time series to investigate in our analysis
 is the \emph{ DAX index} from 1990 to
2009.
 Let $Y(t)$ be the DAX index adjusted close value at day $t$. We define the \emph{DAX index daily return} on day $t$ by
$$
r(t)=\frac{Y(t)-Y(t-1)}{Y(t-1)}
$$
We define the $\alpha$ \emph{re-scaled DAX daily index positive returns} $r(t)^\alpha$, for $r(t)>0$, that we call, after normalization, the $\alpha$ \emph{positive fluctuations}. We define the $\alpha$ \emph{re-scaled DAX daily index negative returns} $(-r(t))^\alpha$, for $r(t)<0$, that we call, after normalization, the $\alpha$  \emph{negative fluctuations}.
We analyze, separately, the $\alpha$ positive and $\alpha$ negative daily fluctuations that can have different statistical and economic natures due, for instance, to the leverage effects (see, for example, \cite{Andersen, Barnhartetal09} and \cite{Pinto1}).
Our aim is to find the values of $\alpha$ that optimize the data collapse of the histogram of the $\alpha$ positive and  $\alpha$ negative
fluctuations to the universal, non-parametric, Bramwell-Holdsofworth-Pinton (BHP) probability density function.
To do it, we apply the Kolmogorov-Smirnov statistic test to the null hypothesis claiming
that the probability distribution of the $\alpha$
fluctuations is equal to the (BHP) distribution.
We observe that the $P$ values of the Kolmogorov-Smirnov test
vary continuously with $\alpha$. The highest $P$ values  $P^{+}=0.19...$ and $P^{-}=0.24...$ of the Kolmogorov-Smirnov test are attained for the values $\alpha^{+}= 0.50...$  and $\alpha^{-}= 0.48...$, respectively, for the positive and negative fluctuations. Hence, the null hypothesis is not rejected for values of $\alpha$ in  small neighborhoods of $\alpha^{+}= 0.50...$  and $\alpha^{-}= 0.48...$.
Then, we show the data collapse of the histograms of the $\alpha^{+}$ positive fluctuations and $\alpha^{-}$ negative fluctuations to the  BHP pdf. Using this data collapse, we do a change of variable that allow us to compute the analytic expressions of the probability density 
 functions $f_{DAX,+}$ and $f_{DAX,-}$
 of the normalized positive and negative  DAX index daily returns
\begin{eqnarray*}
f_{DAX,+}(x)=5.58... x^{-0.50...}f_{BHP}(22.2...x^{0.50...}-1.99...). \\
f_{DAX,-}(x)=4.79...x^{-0.52...}f_{BHP}(20.12....x^{0.48...}-2.01...)
\end{eqnarray*}
in terms of the BHP pdf $f_{BHP}$. 
We exhibit the data collapse of the histogram of the positive and negative returns to 
our proposed theoretical pdf´s $f_{DAX,+}$ and $f_{DAX,-}$.
Similar results are
observed for some other stock indexes, prices of stocks, exchange rates and
commodity prices (see \cite{Gonc, Gond}).
Since the BHP probability density function appears in several other dissimilar phenomena (see, for instance, \cite{bramwellfennelleuphys2002,
DahlstedtJensen2001, Gona, Gong, Pintoetal09}), our result reveals an universal feature of the stock exchange markets.

\section{POSITIVE DAX INDEX DAILY RETURNS}
\label{sec:Positive DAX index daily returns}

Let $T^+$ be the set of all days $t$ with positive returns, i.e.
 $$
 T^+=\{t:r(t)>0\} .
 $$
 Let $n^+=2524$ be the cardinal of the set $T^+$. The \emph{$\alpha$ re-scaled S\&P100 daily index positive returns} are the returns $r(t)^\alpha$ with $t\in T^+$. Since   the total number of observed days is $n=4758$, we obtain that  $n^+/n=0.53$.
 The \emph{mean} $\mu^+_{\alpha}=0.09...$ of the $\alpha$ re-scaled DAX daily index positive returns  is given by
\begin{equation}
\mu^+_{\alpha}=\frac{1}{n^{+}}\sum_{t\in T^+}r(t)^\alpha
 \label{eq2}
\end{equation}
The \emph{standard deviation}  $\sigma^+_{\alpha}=0.045...$ of the $\alpha$ re-scaled DAX daily index positive returns  is given by
\begin{equation}
\sigma^+_{\alpha}=\sqrt{\frac{1}{n^{+}}\sum_{t\in T^+} {r(t)^{2\alpha}} - (\mu^+_{\alpha})^2}
 \label{eq3}
\end{equation}
\noindent
We define the $\alpha$ \emph{positive fluctuations} by
\begin{equation}
r^+_{\alpha}(t) = \frac{r(t)^\alpha - \mu^+_{\alpha}}{\sigma^+_{\alpha}}
 \label{eq6}
\end{equation}
\noindent
for every $t\in T^+$. Hence, the $\alpha$ \emph{positive fluctuations} are the normalized $\alpha$ re-scaled $DAX$ daily index positive returns.
Let $L^+_{\alpha}=-1.90...$ be the \emph{smallest} $\alpha$ positive fluctuation, i.e.
$$
L^+_{\alpha}=\min_{t\in T^+}\{r^+_{\alpha}(t)\}.
$$
Let $R^+_{\alpha}=5.51...$ be the \emph{largest} $\alpha$ positive fluctuation, i.e.
$$
R^+_{\alpha}=\max_{t\in T^+}\{r^+_{\alpha}(t)\}.
$$
We denote by $F_{\alpha,+}$ the \emph{probability distribution of the $\alpha$ positive fluctuations}.
Let the \emph{truncated BHP probability distribution} $F_{BHP,\alpha,+}$ be given by
$$
F_{BHP, \alpha,+}(x)=\frac{F_{BHP}(x)}{F_{BHP}(R^+_{\alpha})-F_{BHP}(L^+_{\alpha})}
$$
where $F_{BHP}$ is the BHP probability distribution.\\

\begin{figure}[htbp!]
\includegraphics[scale=.45]{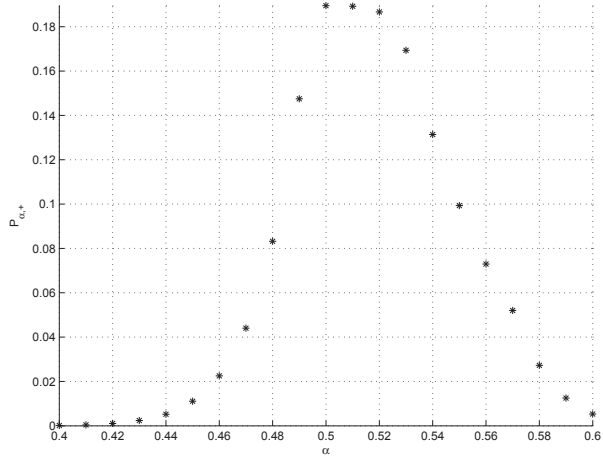}
\caption{The Kolmogorov-Smirnov $P$ value $P_{\alpha,+}$ for values of $\alpha$ in the range $[0.4, 0.6]$. 
}
\label{fig1}
\end{figure}

\noindent We apply the Kolmogorov-Smirnov statistic test to the null hypothesis claiming that the probability distributions $F_{\alpha,+}$ and $F_{BHP,\alpha,+}$ are equal. The Kolmogorov-Smirnov $P$ \emph{value} $P_{\alpha,+}$  is  plotted in Figure \ref{fig1}.
Hence, we observe that $\alpha^+=0.50...$ is the point where the $P$ value $P_{\alpha,+} =0.19...$  attains its maximum.\\

\noindent It is well-known that the Kolmogorov-Smirnov $P$ value $P_{\alpha,+}$  decreases with the distance  
$\left\|F_{\alpha,+}-F_{BHP,\alpha,+}\right\|$
between $F_{\alpha,+}$ and $F_{BHP,\alpha,+}$.\\

\begin{figure}[htbp!]
\includegraphics[scale=.45]{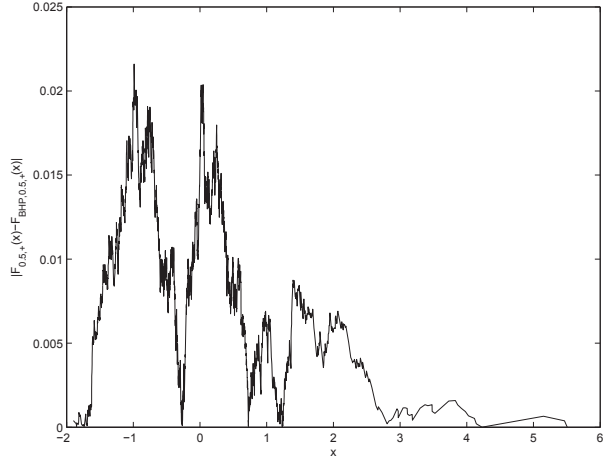}
\caption{The map $D_{0.50,+}(x)=|F_{0.50,+}(x)-F_{BHP,0.50,+}(x)|$.}
 \label{fig2}
\end{figure}

\noindent In Figure \ref{fig2}, we plot $D_{\alpha^+,+}(x)=\left|F_{\alpha^+,+}(x)-F_{BHP,\alpha^+,+}(x)\right|$ and we observe that 
$D_{\alpha^+,+}(x)$ attains its highest values for the $\alpha^+$ positive fluctuations  below or close to the mean of the probability distribution.\\

\begin{figure}[htbp!]
\includegraphics[scale=.45]{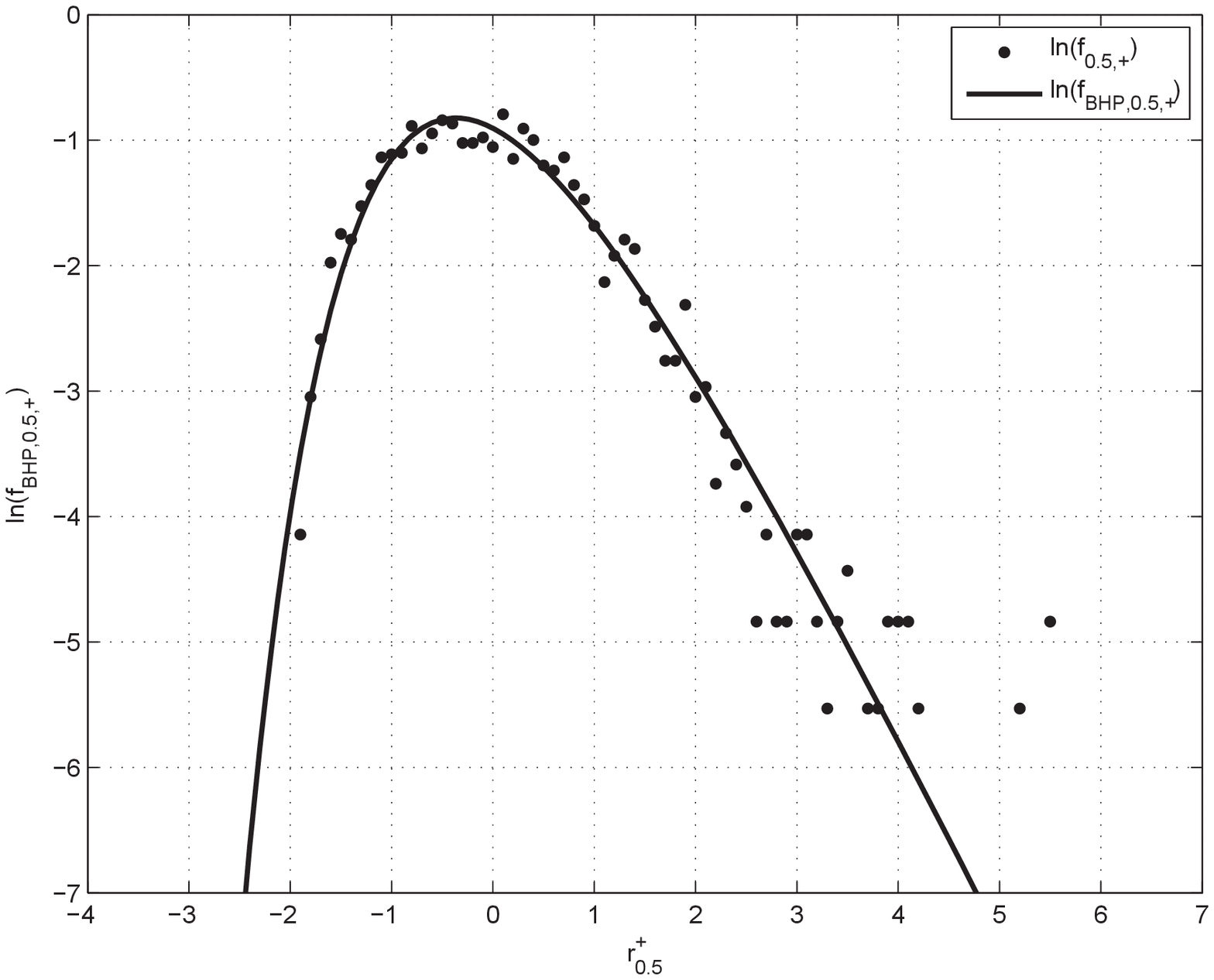}
\caption{The histogram  of the $\alpha^+$ positive fluctuations with the truncated BHP  pdf $f_{BHP,0.50,+}$ on top, in the semi-log scale.}
 \label{fig3}
\end{figure}

\begin{figure}[htbp!]
\includegraphics[scale=.45]{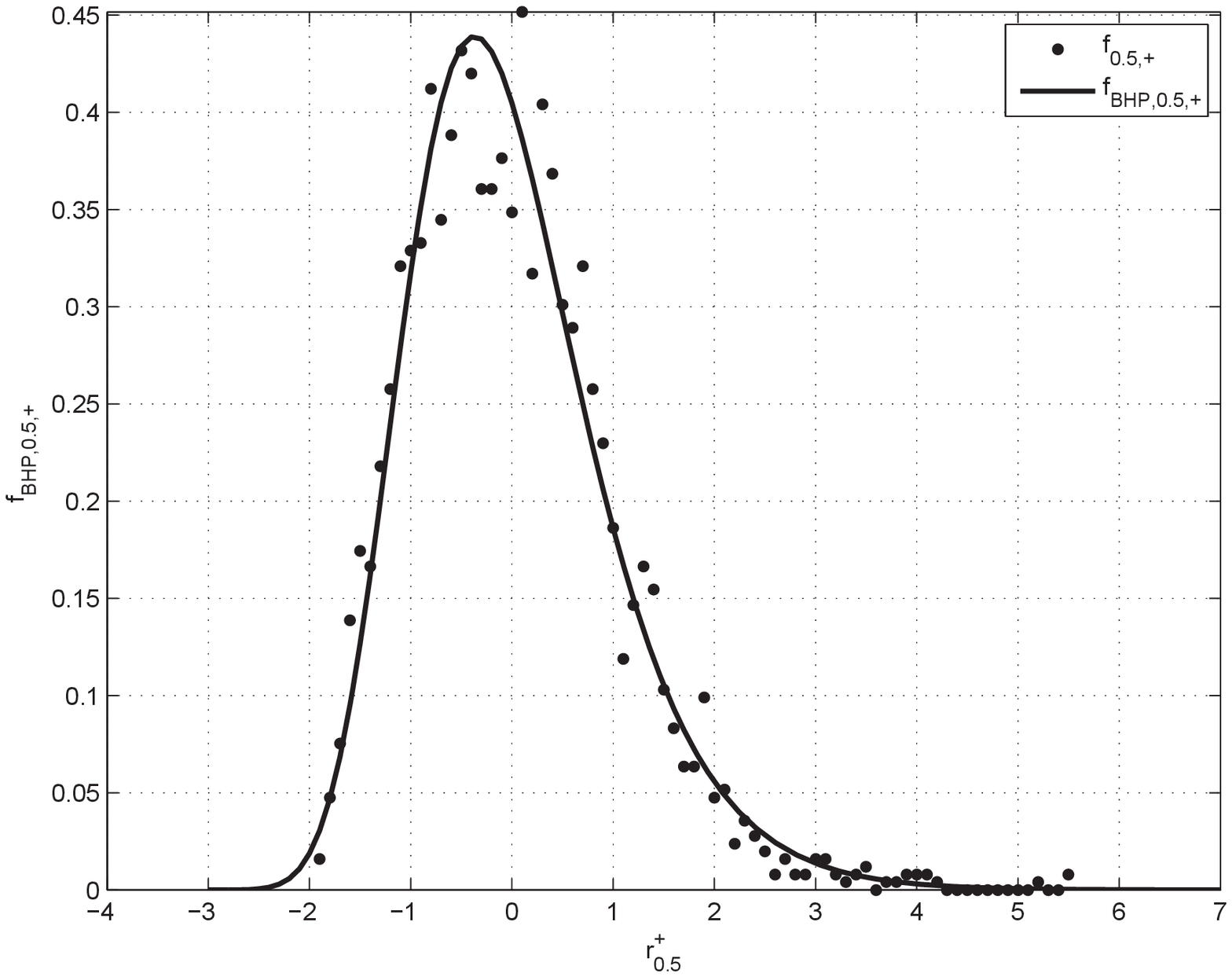}
\caption{The histogram  of the $\alpha^+$ positive fluctuations with the truncated BHP pdf $f_{BHP,0.50,+}$ on top.}
 \label{fig20}
\end{figure}

\noindent In Figures \ref{fig3} and \ref{fig20}, we show the data collapse of the histogram $f_{\alpha^+,+}$ of the $\alpha^+$ positive fluctuations to the  truncated BHP pdf  $f_{BHP,\alpha^+,+}$. \\

\noindent
Assume that the probability distribution of the $\alpha^+$ positive fluctuations $r^+_{\alpha^+}(t)$ is given by $F_{BHP,\alpha^+,+}$, see \cite{Gonb}.
The pdf $f_{DAX,+}$ of the DAX daily index positive returns $r(t)$ is given by
$$
f_{DAX,+}(x)= \frac{\alpha^+ x^{\alpha^+-1}f_{BHP}\left(\left(x^{\alpha^+}-\mu^+_{\alpha^+}\right)/\sigma^+_{\alpha^+}\right)}{\sigma^+_{\alpha^+}\left(F_{BHP}\left(R^+_{\alpha^+}\right)-F_{BHP}\left(L^+_{\alpha^+}\right)\right)}.
$$

\begin{figure}[htbp!]
\includegraphics[scale=.50]{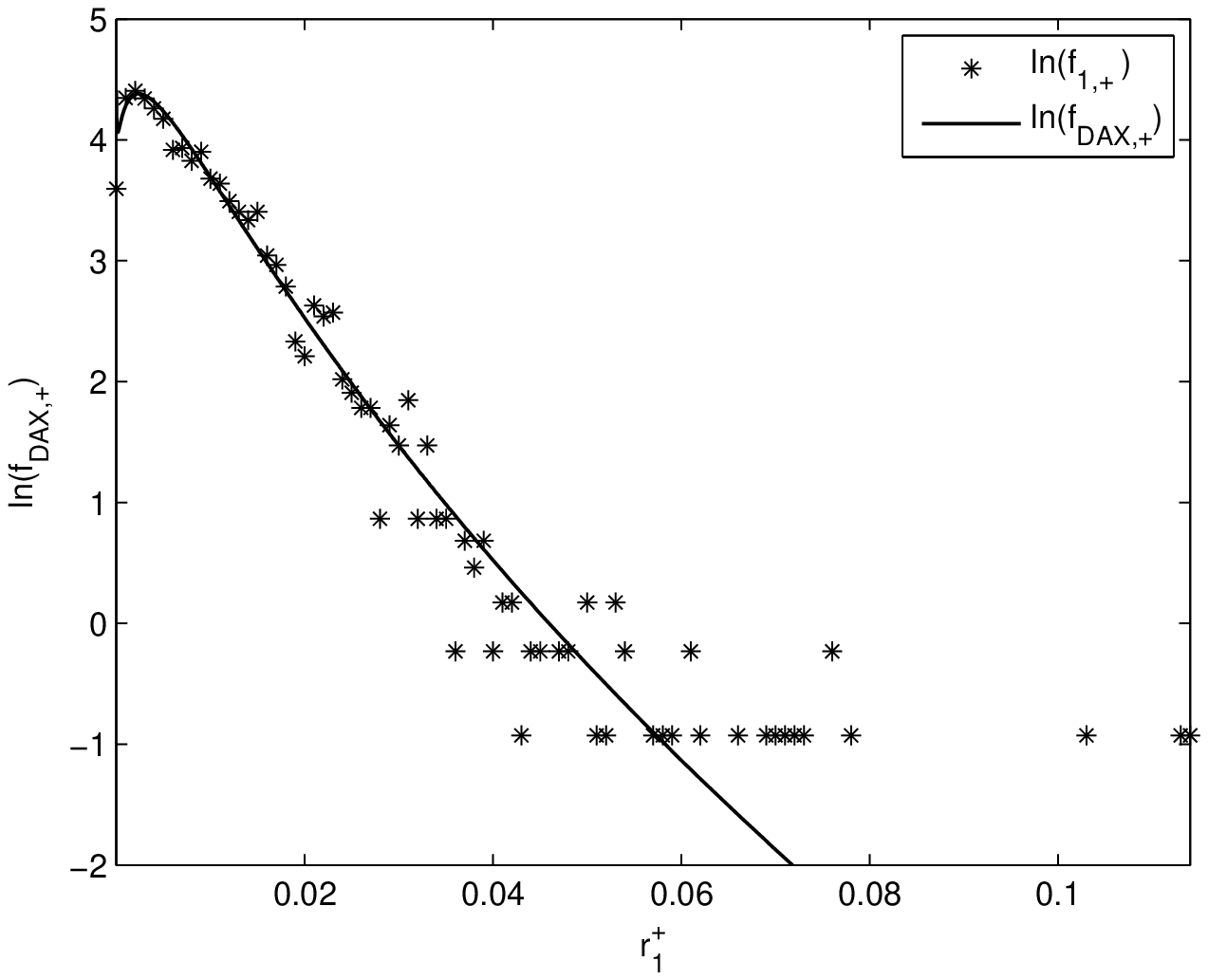}
\caption{The histogram of the fluctuations
 of the positive returns with the pdf $f_{DAX,+}$ on top, in the semi-log scale.}
 \label{fig5}
\end{figure}

\begin{figure}[htbp!]
\includegraphics[scale=.50]{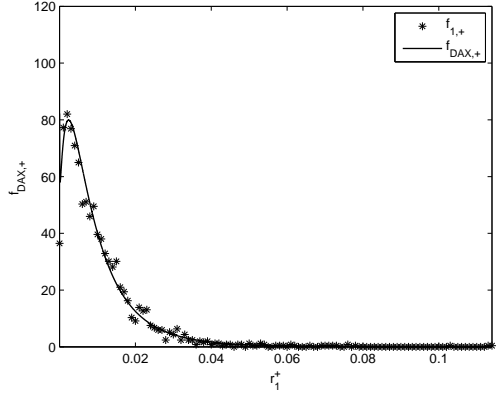}
\caption{The histogram of the fluctuations
 of the positive returns with the pdf $f_{DAX,+}$ on top.}
 \label{fig6}
\end{figure}

\noindent
Hence, taking $\alpha^+=0.50...$, we get
$$
f_{DAX,+}(x)=5.58... x^{-0.50...}f_{BHP}(22.2...x^{0.50...}-1.99...).
$$
\noindent In Figures \ref{fig5} and \ref{fig6} , we show the data collapse of the histogram $f_{1,+}$ of the positive returns to 
our proposed theoretical pdf $f_{DAX,+}$. 

 \noindent
 \noindent 

\section{NEGATIVE DAX INDEX DAILY RETURNS}
\label{sec:Negative DAX index daily returns}

Let $T^-$ be the set of all days $t$ with negative returns, i.e.
 $$
 T^-=\{t:r(t)<0\} .
 $$
 Let $n^-=2221$ be the cardinal of the set $T^-$.  Since   the total number of observed days is $n=4758$, we obtain that  $n^-/n=0.47$.
 The \emph{$\alpha$ re-scaled DAX daily index negative returns} are the returns $(-r(t))^\alpha$ with $t\in T^-$. We note that  $-r(t)$ is positive.
 The \emph{mean} $\mu^-_{\alpha}=0.10...$ of the $\alpha$ re-scaled DAX daily index negative returns  is given by
\begin{equation}
\mu^-_{\alpha}=\frac{1}{n^{-}}\sum_{t\in T^-}(-r(t))^\alpha
 \label{eq2}
\end{equation}
The \emph{standard deviation}  $\sigma^-_{\alpha}=0.050...$ of the $\alpha$ re-scaled DAX daily index negative returns  is given by
\begin{equation}
\sigma^-_{\alpha}=\sqrt{\frac{1}{n^{-}}\sum_{t\in T^-} {(-r(t))^{2\alpha}} - (\mu^-_{\alpha})^2}
 \label{eq3}
\end{equation}
\noindent
We define the $\alpha$ \emph{negative fluctuations} by
\begin{equation}
r^-_{\alpha}(t) = \frac{(-r(t))^\alpha - \mu^-_{\alpha}}{\sigma^-_{\alpha}}
 \label{eq6}
\end{equation}
\noindent
for every $t\in T^-$. Hence, the $\alpha$ \emph{negative fluctuations} are the normalized $\alpha$ re-scaled $DAX$ daily index negative returns.
\noindent Let $L^-_{\alpha}=-1.94...$ be the \emph{smallest} $\alpha$ negative fluctuation, i.e.
$$
L^-_{\alpha}=\min_{t\in T^-}\{r^-_{\alpha}(t)\}.
$$
Let $R^-_{\alpha}=4.45...$ be the \emph{largest} $\alpha$ negative fluctuation, i.e.
$$
R^-_{\alpha}=\max_{t\in T^-}\{r^-_{\alpha}(t)\}.
$$
\noindent We denote by $F_{\alpha,-}$ the \emph{probability distribution of the $\alpha$ negative fluctuations}.
Let the \emph{truncated BHP probability distribution} $F_{BHP,\alpha,-}$ be given by
$$
F_{BHP, \alpha,-}(x)=\frac{F_{BHP}(x)}{F_{BHP}(R^-_{\alpha})-F_{BHP}(L^-_{\alpha})}
$$
where $F_{BHP}$ is the BHP probability distribution.\\

\noindent We apply the Kolmogorov-Smirnov statistic test to the null hypothesis claiming that the probability distributions $F_{\alpha,-}$ and $F_{BHP,\alpha,-}$ are equal.\\
 
\begin{figure}[htbp!]
\includegraphics[scale=.45]{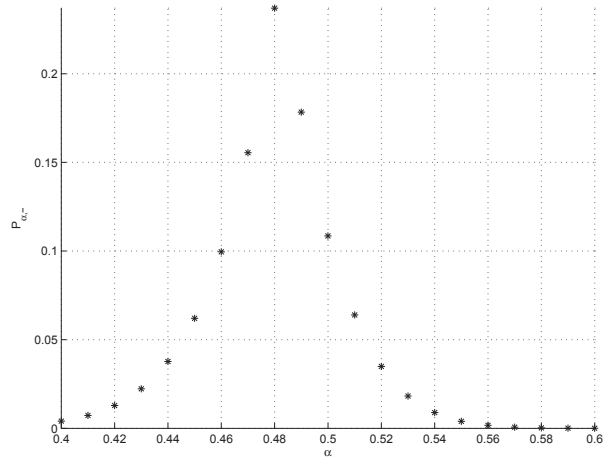}
\caption{The Kolmogorov-Smirnov $P$ value $P_{\alpha,-}$ for values of $\alpha$ in the range $[0.4, 0.6]$.
} \label{fig1x}
\end{figure}

\noindent The Kolmogorov-Smirnov $P$ \emph{value} $P_{\alpha,-}$  is  plotted in Figure \ref{fig1x}. Hence, we observe that $\alpha^-=0.48...$ is the point where the $P$ value $P_{\alpha^-} =0.24...$  attains its maximum.\\

\noindent The Kolmogorov-Smirnov $P$ value $P_{\alpha,-}$  decreases with the distance  
$\left\|F_{\alpha,-}-F_{BHP,\alpha,-}\right\|$
between $F_{\alpha,-}$ and $F_{BHP,\alpha,-}$.\\

\begin{figure}[htbp!]
\includegraphics[scale=.45]{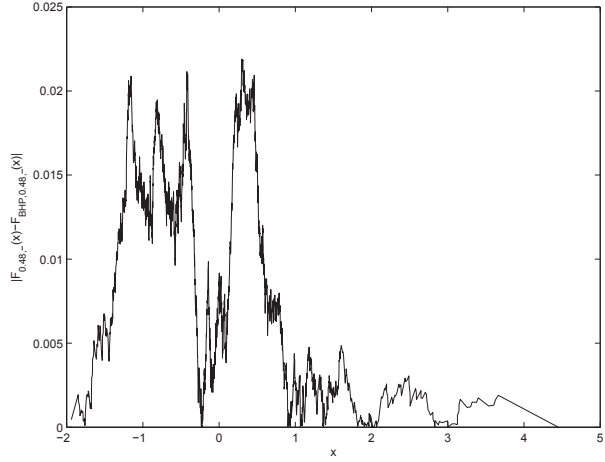}
\caption{The map $D_{0.48,-}(x)=|F_{0.48,-}(x)-F_{BHP,0.48,-}(x)|$.}
\label{fig2x}
\end{figure}

\noindent In Figure \ref{fig2x}, we plot $D_{\alpha^-,-}(x)=\left|F_{\alpha^-,-}(x)-F_{BHP,\alpha^-,-}(x)\right|$ and we observe that 
$D_{\alpha^-,-}(x)$ attains its highest values for the $\alpha^-$ negative fluctuations above the mean of the probability distribution.\\

\begin{figure}[htbp!]
\includegraphics[scale=.45]{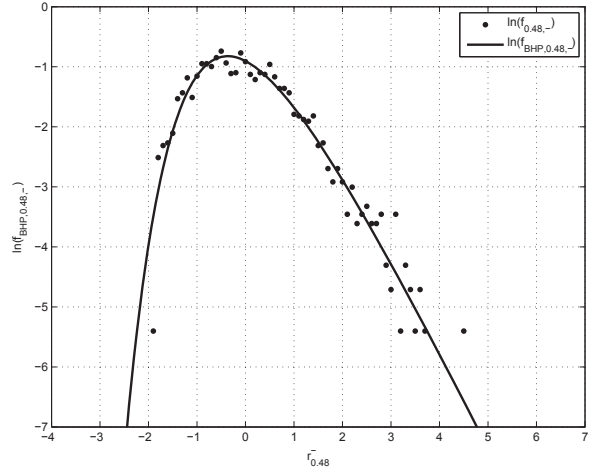}
\caption{The histogram of the $\alpha^-$ negative fluctuations
 with the truncated BHP pdf $f_{BHP,0.48,-}$ on top, in the semi-log scale.}
 \label{fig3x}
\end{figure}

\begin{figure}[htbp!]
\includegraphics[scale=.45]{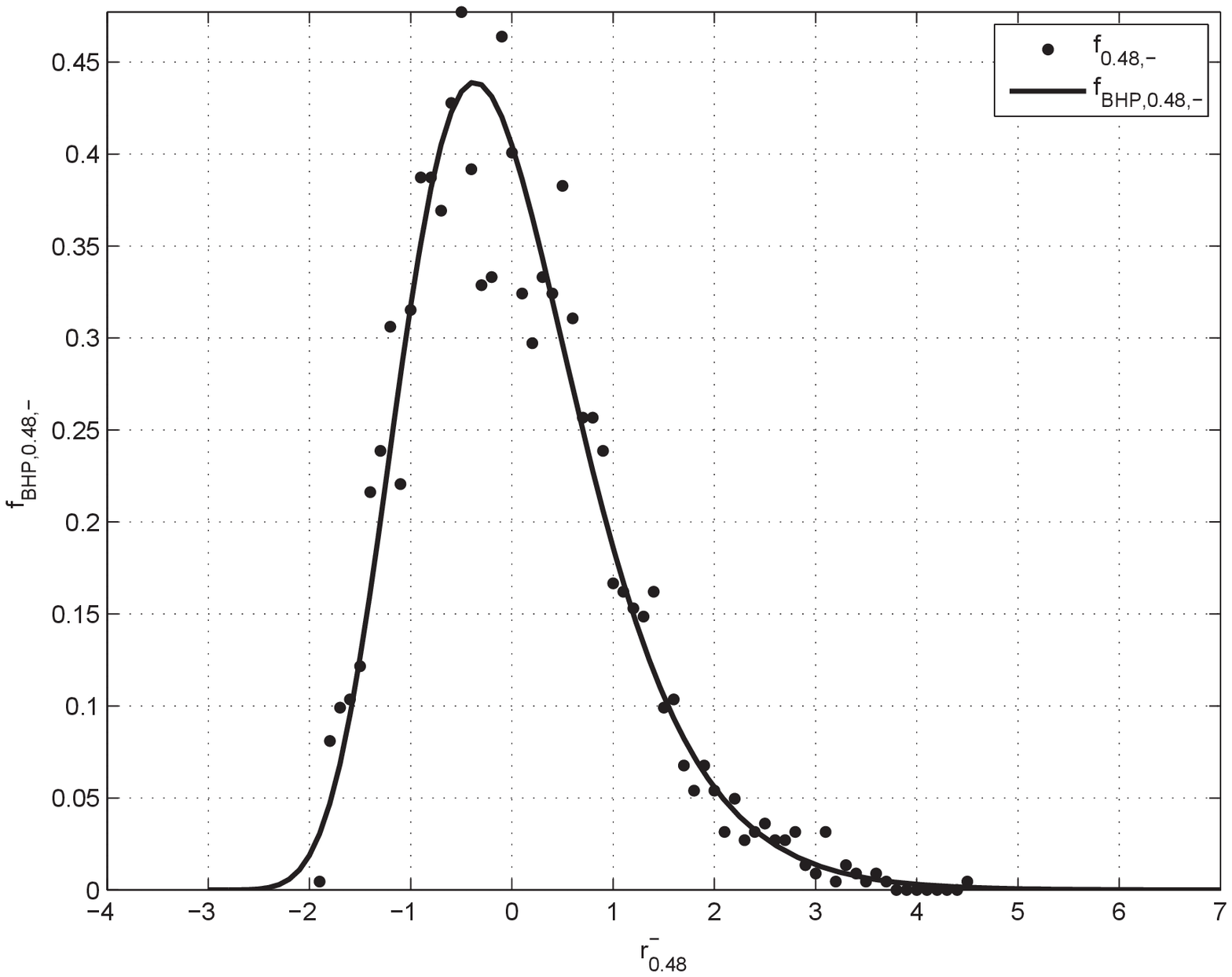}
\caption{The histogram of the $\alpha^-$ negative fluctuations
 with the truncated BHP pdf $f_{BHP,0.48,-}$ on top.}
 \label{fig4x}
\end{figure}

\noindent In Figures \ref{fig3x} and \ref{fig4x}, we show the data collapse of the histogram $f_{\alpha^-,-}$  of the $\alpha^-$ negative fluctuations to the truncated BHP pdf $f_{BHP,\alpha^-,-}$.\\

\noindent Assume that the probability distribution of the $\alpha^-$ negative fluctuations $r^-_{\alpha^-}(t)$ is given by $F_{BHP,\alpha^-,-}$, see \cite{Gonb}.
The pdf $f_{DAX,-}$ of the DAX  daily index (symmetric) negative returns $-r(t)$, with $T \in T^-$,  is given by
$$
f_{DAX,-}(x)=  \frac{\alpha^- x^{\alpha^-1}f_{BHP}\left(\left(x^{\alpha^-}-\mu^-_{\alpha^-}\right)/\sigma^-_{\alpha^-}\right)}{\sigma^-_{\alpha^-}\left(F_{BHP}\left(R^-_{\alpha^-}\right)-F_{BHP}\left(L^-_{\alpha^-}\right)\right)}.
$$

\noindent
Hence, taking $\alpha^-=0.49...$, we get
$$
f_{DAX,-}(x)=4.79...x^{-0.52...}f_{BHP}(20.12....x^{0.48...}-2.01...)
$$

\begin{figure}[htbp!]
\includegraphics[scale=.50]{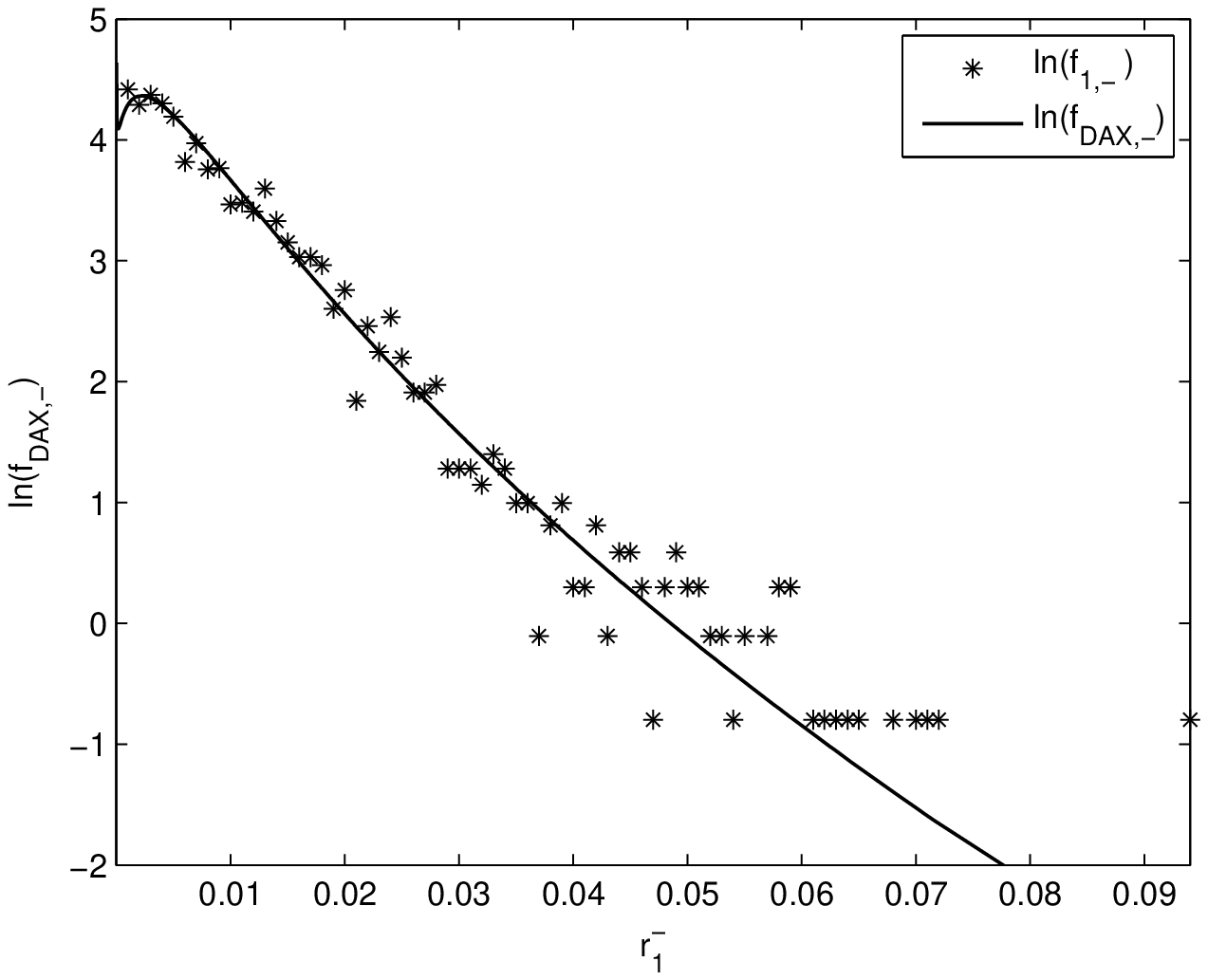}
\caption{The histogram of the negative returns with the pdf $f_{DAX,-}$ on top, in the semi-log scale.}
 \label{fig5x}
\end{figure}

\begin{figure}[htbp!]
\includegraphics[scale=.50]{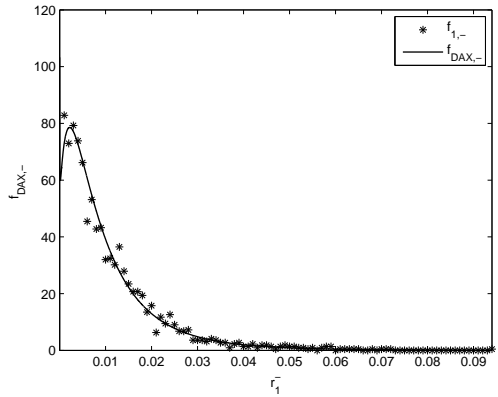}
\caption{The histogram of the negative returns with the pdf $f_{DAX,-}$ on top.}
 \label{fig6x}
\end{figure}

\noindent In Figures \ref{fig5x} and \ref{fig6x}, we show the data collapse of the histogram $f_{1,-}$ of the negative returns to 
our proposed theoretical pdf $f_{DAX,-}$.

\newpage
\section{CONCLUSIONS}
We used the Kolmogorov-Smirnov statistical test to compare the histogram of the $\alpha$ positive fluctuations and $\alpha$ negative
fluctuations with the universal, non-parametric, Bramwell-Holdsworth-Pinton (BHP) probability distribution.
We found that the parameters $\alpha^{+}= 0.50...$  and $\alpha^{-}= 0.48...$ for the positive and negative fluctuations, respectively, optimize the $P$ value of the Kolmogorov-Smirnov test. We obtained that the respective $P$ values of the
Kolmogorov-Smirnov statistical test are $P^{+}=0.19...$ and $P^{-}=0.23...$. Hence, the null hypothesis was not rejected.
The fact that $\alpha^+$ is different from $\alpha^-$ can be do to leverage effects.
We presented the data collapse of the corresponding fluctuations histograms to the BHP pdf.
Furthermore, we computed the analytic expression of the probability distributions $F_{DAX,+}$ and $F_{DAX,-}$ of the normalized DAX index daily positive and negative returns in terms of the BHP pdf. We showed the data collapse of the histogram of the positive and negative returns to 
our proposed theoretical pdfs $f_{DAX,+}$ and $f_{DAX,-}$. The results obtained in daily returns also apply to other periodicities, such as weekly and monthly returns as well as intraday values.

In \cite{Gonc, Gonb, science},  it is found the 
data collapses of the histograms of some other stock indexes, prices of stocks, exchange rates and
commodity prices to the BHP pdf and in \cite{s6} for energy sources.

Bramwell, Holdsworth and Pinton \cite{BHP1998} found the
probability distribution of the fluctuations of the total magnetization,
in the strong coupling (low temperature) regime, for a
two-dimensional spin model (2dXY) using the spin wave
approximation. From a statistical physics point of view,
one can think that the stock prices form a non-equilibrium system
\cite{Chowdhury, Gopikrishnanetal98, LilloMan01, Plerouetal99}.
Hence, the results presented here lead to a construction of a new qualitative and quantitative
econophysics model for the stock market based in the two-dimensional spin
model (2dXY) at criticality (see \cite{Gond}).

\section*{Acknowledgments}
We thank Peter Holdsworth and Henrik Jensen for
showing us the relevance of the BHP
distribution.This work was presented in PODE09, EURO XXIII, Encontro Ci\^encia 2009 and ICDEA2009.
We thank LIAAD-INESC Porto LA, Calouste Gulbenkian Foundation, PRODYN-ESF, POCTI and POSI by FCT and Minist\'erio da Ci\^encia e da Tecnologia, and the FCT Pluriannual Funding Program of the LIAAD-INESC Porto LA. Part of this research was developed during a visit by the authors to the IHES, CUNY, IMPA, MSRI, SUNY, Isaac Newton Institute and University of Warwick. We thank them for their hospitality.

\appendix
\section{BRAMWELL-HOLDSWORTH-PINTON PROBABILITY DISTRIBUTION}
The universal nonparametric BHP pdf was discovered by Bramwell,
Holdsworth and Pinton \cite{BHP1998}. The \emph{BHP probability density function (pdf)} is
given by
\begin{eqnarray}
             &\!&f_{BHP}(\mu)=\int_{-\infty}^{\infty}\frac{dx}{2\pi}
\sqrt{\frac{1}{2N^2}\sum_{k=1}^{N-1}\frac{1}{\lambda_k^2}}
e^{ix\mu\sqrt{\frac{1}{2N^2}\sum_{k=1}^{N-1}\frac{1}{\lambda_k^2}}}\nonumber\\\!
&\!&
.e^{-\sum_{k=1}^{N-1}\left[\frac{ix}{2N}\frac{1}{\lambda_k}-\frac{i}{2}
\mbox{arctan}\left(\frac{x}{N\lambda_k}\right)\right]}.e^{-\sum_{k=1}^{N-1}\left[\frac{1}{4}\mbox{ln}{\left(1+\frac{x^2}{N^2\lambda_k^2}\right)}\right]}\nonumber\\
      \label{eq1}
\end{eqnarray}
\noindent where the $\{\lambda_k\}_{k=1}^L$ are the eigenvalues, as
determined in \cite{Bramwelletal2001}, of the adjacency matrix. It
follows, from the formula of the BHP pdf, that the asymptotic values
for large deviations, below and above the mean, are exponential and
double exponential, respectively (in this article, we use the
approximation of the BHP pdf obtained by taking $L=10$ and $N=L^2$
in equation (\ref{eq1})). As we can see, the BHP distribution does
not have any parameter (except the mean that is normalize to 0 and
the standard deviation that is normalized to 1) and it is universal,
in the sense that appears in several physical phenomena. For
instance, the universal nonparametric BHP distribution is a good
model to explain the fluctuations of order parameters in theoretical
examples such as, models of self-organized criticality, equilibrium
critical behavior, percolation phenomena (see \cite{BHP1998}), the
Sneppen model (see \cite{BHP1998} and \cite{DahlstedtJensen2001}),
and auto-ignition fire models (see \cite{SinharayBordaJensen2001}).
The universal nonparametric BHP distribution is, also, an
explanatory model for fluctuations of several phenomenon such as,
width power in steady state systems (see \cite{BHP1998}),
fluctuations in river heights and flow (see \cite{Bramwelletal2001, Gona, Gonf}), for the plasma density
fluctuations and electrostatic turbulent fluxes measured at the
scrape-off layer of the Alcator C-mod Tokamaks (see
\cite{VanMilligen05}) and for Wolf's sunspot numbers
 fluctuations (see \cite{Gong}).
 
\bibliographystyle{apsrmp}


\begin{thebibliography}{42}

\bibitem[Andersen(2004) Andersen, Bollerslev, Frederiksen, and Nielse]{Andersen}
Andersen, T.G., Bollerslev, T., Frederiksen, P. and Nielse, M., 2004,
Continuos-Time Models, Realized Volatilities and Testable Distributional Implications for Daily Stock Returns
{\it Preprint}.

\bibitem [Barnhart(2009) Barnhart and Giannetti] {Barnhartetal09}
Barnhart, S. W. \& Giannetti, A., 2009, Negative earnings, positive earnings and stock return predictability: An empirical examination of market timing
{\it Journal of Empirical Finance} {\bf 16} 70-86.


\bibitem [Bramwell(1998)Bramwell, Holdsworth and Pinton]{BHP1998}
Bramwell, S.T., Holdsworth, P.C.W., \&  Pinton, J.F., 1998,{\it Nature}, {\bf 396} 552-554.

\bibitem[Bramwell(2002)Bramwell, Holdsworth and Pinton]{bramwellfennelleuphys2002}
 Bramwell, S.T., Fennell, T., Holdsworth, P.C.W.,\& Portelli, B., 2002, Universal Fluctuations of the Danube Water Level: a Link with Turbulence, Criticality and Company Growth, {\it Europhysics Letters  }{\bf 57} 310.


\bibitem [Bramwell(2001)Bramwell, Fortin, Holdsworth, Pinton, Portelli and Sellitto]{Bramwelletal2001}
Bramwell, S.T., Fortin, J.Y., Holdsworth, P.C.W., Peysson, S., Pinton, J.F., Portelli, B. \& Sellitto, M., 2001,
Magnetic Fluctuations in the classical XY model: the origin of an exponential tail in a complex system, {\it Phys. Rev  }{\bf E 63} 041106.



\bibitem [Chowdhury(1999)Chowdhury and Stauffer]{Chowdhury}
Chowdhury, D. and Stauffer, D., 1991, A generalized spin model of financial markets
{\it Eur. Phys. J.} {\bf B8} 477-482.

\bibitem[Dahlstedt(2001)Dahlstedt and Jensen]{DahlstedtJensen2001}
Dahlstedt, K., \& Jensen, H.J., 2001, Universal fluctuations and extreme-value statistics, {\it J. Phys. A: Math. Gen. }{\bf 34} 11193-11200.


\bibitem [Peixoto(2001) Peixoto, Pinto and Rand]{science}
Dynamics, Games and Science., 2010, 
Eds: M. Peixoto, A. A. Pinto and D. A.  Rand.
Proceedings in Mathematics series, Springer-Verlag.

\bibitem [Gabaix(2001) Gabaix, Parameswaran, Plerou, Stanley]{Gabaixetal03}
Gabaix, X., Parameswaran, G., Plerou, V. \& Stanley, E., 2003, A theory of power-law distributions in financial markets {\it Nature} {\bf 423} 267-270.

\bibitem[Gon\c calves(2009a)  Gon\c calves, Ferreira and Pinto]{Gonb}
Gon\c calves, R., Ferreira, H. and Pinto, A. A., 2009a, 
Universality in the Stock Exchange Market {\it Journal of Difference Equations and Applications} (accepted).

\bibitem [Gon\c calves(2010a)  Gon\c calves, Ferreira and Pinto] {s6}
Gon\c calves, R., Ferreira, H. and Pinto, A. A., 2010a, {\em Universality in energy sources},
 IAEE (International Association for Energy economics) International Conference (accepted).

\bibitem[Gon\c calves(2010b)  Gon\c calves, Ferreira and Pinto]{Gonc}
Gon\c calves, R., Ferreira, H. and  Pinto, A. A., 2010b, Universal fluctuations of the Dow Jones (submitted).

\bibitem [Gon\c calves(2010c)  Gon\c calves, Ferreira and Pinto]{Gond}
Gon\c calves, R., Ferreira, H. and Pinto, A. A., 2010c, A qualitative and quantitative Econophysics stock market model (submitted).


\bibitem [Gon\c calves(2009b)  Gon\c calves, Ferreira, Pinto and Stollenwerk]{Gona}
Gon\c calves, R., Ferreira, H., Pinto, A. A. and Stollenwerk, N., 2009b, Universality in nonlinear prediction of complex systems.
Special issue in honor of Saber Elaydi. {\it Journal of Difference Equations and Applications} {\bf 15}, Issue 11 \& 12, 1067-1076.


\bibitem [Gon\c calves(2009c) Gon\c calves and Pinto]{Gonf}
Gon\c calves, R., and Pinto, A. A., 2009c, Negro and Danube are mirror rivers. Special issue Dynamics \& Applications in honor of Mauricio Peixoto and David Rand. {\it Journal of Difference Equations and Applications}.


\bibitem[Gon\c calves(2009d)  Gon\c calves, Pinto and Stollenwerk]{Gong}
Gon\c calves, R. Pinto, A. A., Stollenwerk, N., 2009d, Cycles and universality in sunspot numbers fluctuations {\it The Astrophysical Journal} {\bf 691} 1583-1586.

       
\bibitem [Gopikrishnan(1998)  Gopikrishnan, Meyer, Amaral and Stanley]{Gopikrishnanetal98}
Gopikrishnan, P., Meyer, M., Amaral, L. \& Stanley, H., 1998, Inverse cubic law for the distribution of stock price variation {\it The European Physical Journal B} {\bf 3} 139-140.


\bibitem [Lillo(2001)Lillo, F. and Mantegna, R.]{LilloMan01}
Lillo, F. and Mantegna, R., 2001, Ensemble Properties of securities traded in the Nasdaq market
{\it Physica A} {\bf 299} 161-167 (2001).



\bibitem[Mantegna(2001) Mantegna and Stanley]{ManStan95}
Mantegna, R. \& Stanley, E., 2001, Scaling behaviour in the dynamics of a economic index
 {\it Nature} {\bf 376} 46-49.


\bibitem[Pinto(2010) Pinto]{Pintoetal09}
Pinto, A. A., 2010, Game theory and Duopoly Models {\it Interdisciplinary Applied Mathematics}, Springer-Verlag.
        
\bibitem[Pinto(2009) Pinto, Rand and Ferreira]{Pinto1}        
Pinto, A. A., Rand, D. A. and Ferreira, F.,  2009, 
Fine Structures of Hyperbolic Diffeomorphisms. 
{\it Springer-Verlag Monograph}.

\bibitem [Plerou(1999) Plerou, Amaral, Gopikrishnan, Meyer and Stanley]{Plerouetal99}
Plerou, V., Amaral, L., Gopikrishnan, P., Meyer, M. \& Stanley, E., 1999, Universal and Nonuniversal Properties of Cross Correlations in Financial Time Series {\it Physical Review Letters} {\bf 83} 7 1471-1474.

\bibitem [Sinha-Ray(2001) Sinha-Ray, Borda de \'Agua and Jensen]{SinharayBordaJensen2001}
Sinha-Ray, P., Borda de \'Agua, L. \& Jensen, H.J., 2001, Threshold dynamics, multifractality and universal fluctuations in the SOC forest fire: facets of an auto-ignition model {\it Physica }{\bf D 157}, 186--196.


\bibitem [Van Milligen(2005) Van Milligen, S\'anchez, Carreras,  Lynch,
        LaBombard, Pedrosa,  Hidalgo, Gon\c calves and Balb\' in]{VanMilligen05}
Van Milligen, B. Ph.,  S\'anchez, R., Carreras, B. A., Lynch, V. E., LaBombard, B., Pedrosa,  M. A., Hidalgo, C., Gon\c calves, B. \&  Balb\' in, R., 2005, {\it Physics of plasmas} {\bf 12} 05207.

\end{thebibliography}

\end{document}